\documentclass[twocolumn,floatfix,prl]{revtex4}

\usepackage[]{amsmath}
\usepackage[psamsfonts]{amsfonts}
\usepackage[psamsfonts]{amssymb}

\usepackage{graphicx}
\usepackage{bm}

\usepackage[usenames,dvipsnames]{color}

\newcommand{\newtext}[1]{\textcolor{blue}{#1}}

\newcommand{\urusi}{URu$_{2}$Si$_{2}$}

\begin{document}

\title{Itinerant magnetic multipole moments of rank five, triakontadipoles, as the hidden order in URu$_{2}$Si$_{2}$}

\author{Francesco Cricchio}
\affiliation{Department of Physics and Materials Science, Uppsala University, Box 530, SE-75121 Uppsala, Sweden}  
\author{Fredrik Bultmark}
\affiliation{Department of Physics and Materials Science, Uppsala University, Box 530, SE-75121 Uppsala, Sweden}
\author{Oscar Gr\aa n\"as}
\affiliation{Department of Physics and Materials Science, Uppsala University, Box 530, SE-75121 Uppsala, Sweden} 
\author{Lars Nordstr\"om}
\affiliation{Department of Physics and Materials Science, Uppsala University, Box 530, SE-75121 Uppsala, Sweden}

\begin{abstract}
{A broken symmetry ground state without any magnetic moments has been calculated by means of local-density-approximation to density functional theory plus a local exchange term, the so-called LDA+$U$ approach, for \urusi. The solution is analysed in terms of a multipole tensor expansion of the itinerant density matrix and is found to be a non-trivial magnetic multipole.
Analysis and further calculations show that this type of multipole enters naturally in time reversal breaking in presence of large effective spin-orbit coupling and co-exists with magnetic moments for most magnetic actinides.
}
\end{abstract}

\maketitle

The magnetism involving the 5$f$ states of the actinides shows many exotic behaviours. One of most enigmatic is the tiny magnetic moment \cite{Palstra:1985p4395,Broholm:1991p4795} observed in 
tetragonal URu$_{2}$Si$_{2}$ below a transition temperature of 17 K. It is established that the true ordering parameter driving the second order transition is hidden, i.e.\ not observable by standard techniques.
A huge experimental effort has been made to uncover the true nature of this hidden-order. 
This has led to a large progress in determining the relevant phase diagram of {\urusi}, but
the order stay concealed.
Under pressure the ordering temperature increases slightly, but more significantly for pressures above 0.5 GPa  there is a phase transition to an anti-ferromagnetic (AF) state with magnetic moments of 0.3 $\mu_\mathrm{B}$\cite{Amitsuka:1999p540}. The critical pressure increases with temperature and seems to reach a bicritical point around 1.3 GPa. 
The two phases HO and AF have some properties in common. They both show large anomalies in 
thermodynamical\cite{Hassinger:2008p2528} and transport quantities\cite{Sharma:2006p12198} at the critical temperature, signalling a large Fermi surface nesting.
In fact these quantities\cite{Hassinger:2008p2528} as well as the Fermi surface geometry determined by  de Haas--van Alphen \cite{Nakashima:2003p5427} show only smooth variations across the phase transition. Since there is a sharp transition between the two phases, and since the tiny moments observed in the HO phase are argued to be caused by
 sample dependent extrinsic effects, the order parameter in the HO can in principle be of completely different symmetry than in the AF phase. Especially, the HO might be time reversal (TR) symmetric although the AF clearly is not.
On the other hand, the inflection point observed in the induced moment of an external magnetic field, indicates that there is an  ``adiabatic continuity'' between the two phases and hence that HO is TR odd\cite{Jo:2007p4183}.
In summary, several crucial aspects of the HO are still under debate: whether the HO is of even or odd TR symmetry, whether it is of localised or itinerant character, and whether the HO and AF phases are related or of different origin.
Many explanations, more or less exotic, have been suggested for this ``hidden order'' (HO), some of localised nature e.g.~quadrupoles\cite{Santini:1994p6460} or octupoles\cite{Kiss:2005p359}, and some of itinerant nature as e.g.\ spin nematics\cite{Barzykin:1993p738}, orbital currents\cite{Chandra:2002p13734}, helicity order\cite{Varma:2006p365}, or fluctuating moments\cite{Elgazzar:2008p2090}, but none {of them} have been 
consistent with all experimental observation {in a satisfactory way}.

In this Letter we find that \newtext{a} non-trivial magnetic triakontadipole (rank five) moments constitue the HO parameter in 
URu$_{2}$Si$_{2}$. 
In fact we observe that these multipole moments play an important role in most magnetically ordered materials with a large effective spin-orbit coupling {(SO)}.
Our calculations which treat the 5$f$ states as itinerant with large Coulomb interaction show that {what makes URu$_{2}$Si$_{2}$ unique is} the very large triakontadipole moment {which} forces both spin and orbital dipole moments to vanish. 
In this material the Fermi surface nesting is found to be important for the stabilization of the calculated staggered multipole moments.
With decreased in-plane lattice constant the dipoles increase giving rise to ordinary magnetic moments in accordance with pressure experiments\cite{Amitsuka:1999p540} as the magnitude of the magnetic multipole decreases slightly. 

The driving mechanism for magnetic ordering in conventional magnets based on 3$d$ transitional metal elements, is known to be the Stoner-like exchange, which is for instance included in local density approximation (LDA) {to} the spin density functional theory (SDFT) approach\cite{Kubler}. This gives in most cases a very good description  of magnetic moments and magnetic ordering, whether ferromagnetic, anti-ferromagnetic or non-commensurate spin density waves.
When relativistic effects, especially the  SO coupling, start to play an important role as in the actinides, the Stoner-like spin-polarisation is not sufficient anymore. It tends to over-estimate the spin contribution to the magnetic moments while simultaneously it drastically under-estimates the orbital contribution. This is known to be remedied by {adding} a more general exchange terms, that includes for instance an orbital polarisation term\cite{Brooks,Eriksson}. 
A general form of exchange interactions within an atomic open shell is the screened Hartree-Fock term.
This term is fused with the conventional SDFT formalism in the so-called  LDA+$U$ method\cite{Liechtenstein:1995p15,Solovyev:2005p134}.
Recently we have shown that with this type of exchange, the {SO} coupling can be strongly enhanced, i.e.\ the dominating tensor moment contribution {comes} from the multipole tensor $\mathbf{w}^{110}$. In the case of $\delta$-Pu this was shown to quench the spin-polarisation leading to a TR even state\cite{Cricchio:2008p1260} in accordance with experiments.
The analysis of the ground state was achieved by a general multi-pole expansion of the exchange energy,
$E_\mathrm{X}=\sum_{kpr} K_{kpr} \mathbf{w}^{kpr}\cdot\mathbf{w}^{kpr}$, where $ \mathbf{w}^{kpr}$ is the multipole tensor moment for the channel $kpr$ and $K_{kpr}$ is the corresponding exchange parameter 
These $\mathbf{w}^{kpr}$ tensors correspond to density multipoles for even $k$ even and current multipoles for odd. 
The index $p=0$ describes the charge density and current while $p=1$ describes the spin density and curent.
The orbital ($k$) and spin ($p$) indices are coupled to the index $r$.
Hence $\mathbf{w}^{110}$ corresponds to a scalar product of spin and orbital current $s\cdot\ell$, while e.g.\ 
$\mathbf{w}^{011}$ and $\mathbf{w}^{101}$ are proportional to spin and orbital moments.

In this study we apply  the LDA+$U$ approach {and its multipole decomposition} to a few different magnetic uranium systems, 
three supposedly normal systems;
UAs, US and USb in the NaCl-structure, and {\urusi} with its enigmatic HO in the tetragonal ThCr$_{2}$Si$_{2}$-structure. 
The calculations are set up according to the observed crystal structure\cite{Lander:1995p1293,Palstra:1985p4395} as well as the observed magnetic ordering wave vectors; a single $q=(00\frac{1}{2})$,
$q=(000)$ and a triple $q=(00\frac{1}{2})$, respectively for the three ``normal'' compounds.
The calculations where done in the APW+{\em lo}\cite{Singh} method together with the LDA+$U$ scheme as implemented\cite{Cricchio:2008p1260,Long} in the code {\sc Elk}\cite{elk}.  The basis cut off $R_\mathrm{U}G_\mathrm{max}$, where $R_\mathrm{U}$ is the muffin tin radius of uranium, was set to 9.5. The integration over the BZ zone was performed with a $12\times12\times12$ $k$-points for the fcc cell of the ferromagnets UAs and US, with $8\times8\times8$ $k$-points for the simple cubic cell used in the triple $q$ ordering of USb , and with $18\times18\times10$ for the simple tetragonal cell of the single $q$ ordering of {\urusi}.

\begin{figure}
  \centering
  \includegraphics[width=0.6\columnwidth]{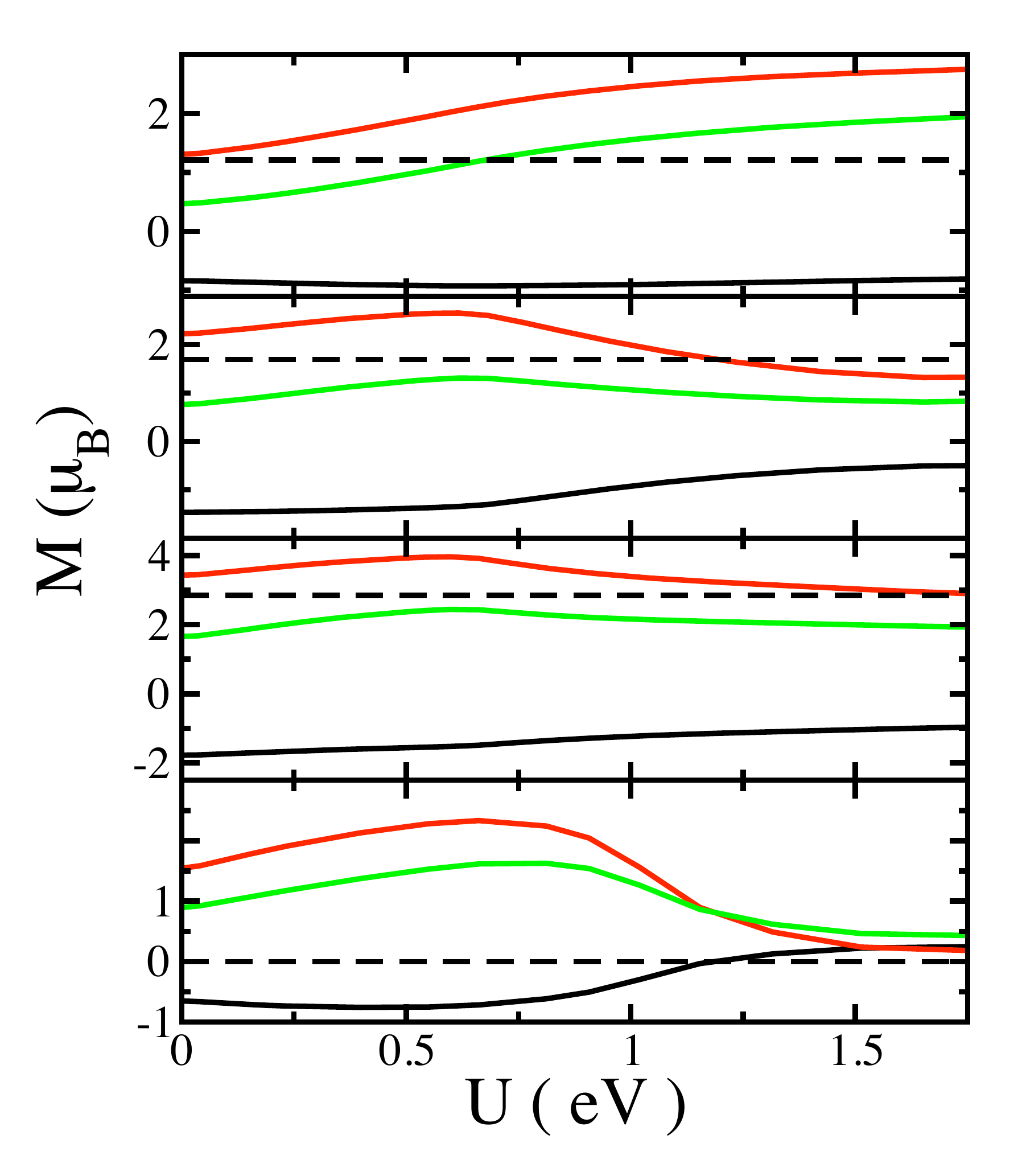}
    \includegraphics[width=0.8\columnwidth]{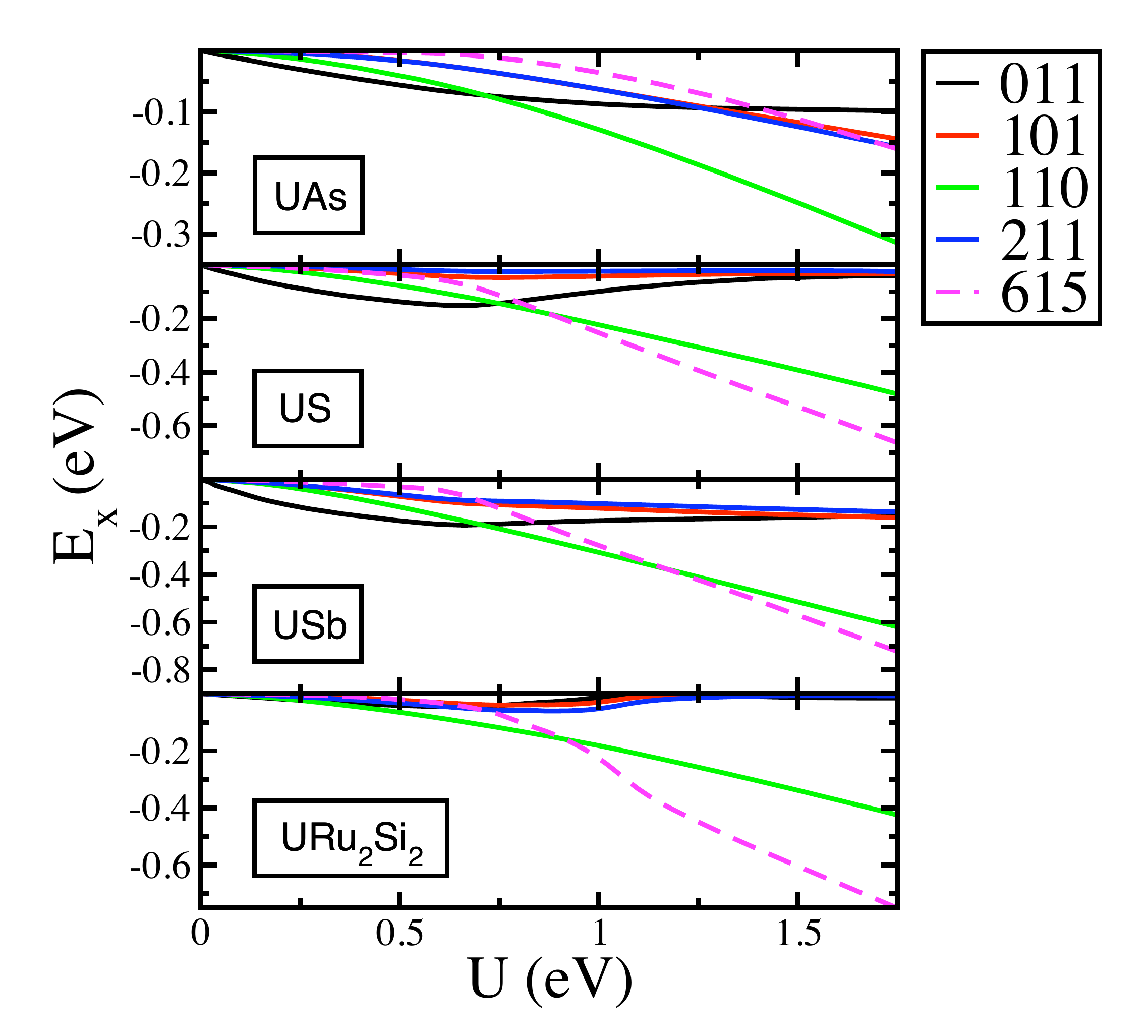}
  \caption{At top: The calculated spin (black), orbital (red) and total (green) magnetic 5$f$ moments as a function of $U$, for various metallic and magnetic uranium compounds. $U=0$ correspond to the LDA-limit. The experimental magnetic moment is given by the dashed line.
At bottom:  The contributions to the exchange energy from the relevant polarisation channels $kpr$ as a function of $U$.  \label{fig:exengy}}
\end{figure}

The calculated magnetic moments are displayed in Fig.\ \ref{fig:exengy} as a function of the parameter $U$. We notice that we have a {good} agreement with the experimental moments\cite{Lander:1995p1293} for $U$ in the range of $0.6$--$1.0$ eV, which is the range of reasonable values for the 5$f$ states of U. These moments are also of the same magnitude as in other ``beyond-LDA'' calculations\cite{Brooks:2004p12779,Solovyev:2005p134}. The calculated spin and orbital moments (SM and OM) show slightly different behaviour for the different uranium compounds, for UAs  the OM increases monotonously with $U$ and the SM is constant while for US and USb the OM has a maximum value around 0.6 eV and the SM monotonously decrease. The last behaviour is somewhat counter-intuitive -- the SM  decrease with increased exchange energy. Here we should note that we have adopted a strategy where the different Slater parameters are screened equivalently \cite{MRNorman:1995p140,Long}. This has advantage that the limit of vanishing $U$ corresponds to an ordinary LDA calculation.

In order to understand why the spin-polarisation decreases with $U$, we display
 the leading tensor contributions to the exchange energy\cite{{Cricchio:2008p1260},Long} 
as a function of the screened $U$-parameter in Fig.\ \ref{fig:exengy}. We can see that the spin-orbit coupling like multipole $\mathbf{w}^{110}$ has a large contribution to the total exchange energy. This leads in turn to that the spin-polarisation decrease in importance compared to the LDA-limit, $U=0$. 
Although the finite $U$ give rise to an orbital polarisation, the most striking effect is that the a non-trivial high order multi-pole dominates, the $\mathbf{w}^{615}$ tensor, and that it plays a large role for all the uranium compounds.
As given by its $kpr$ indices this contribution arises from a multipole of the spin magnetization density.

\begin{figure}
  \centering
  \includegraphics[width=\columnwidth]{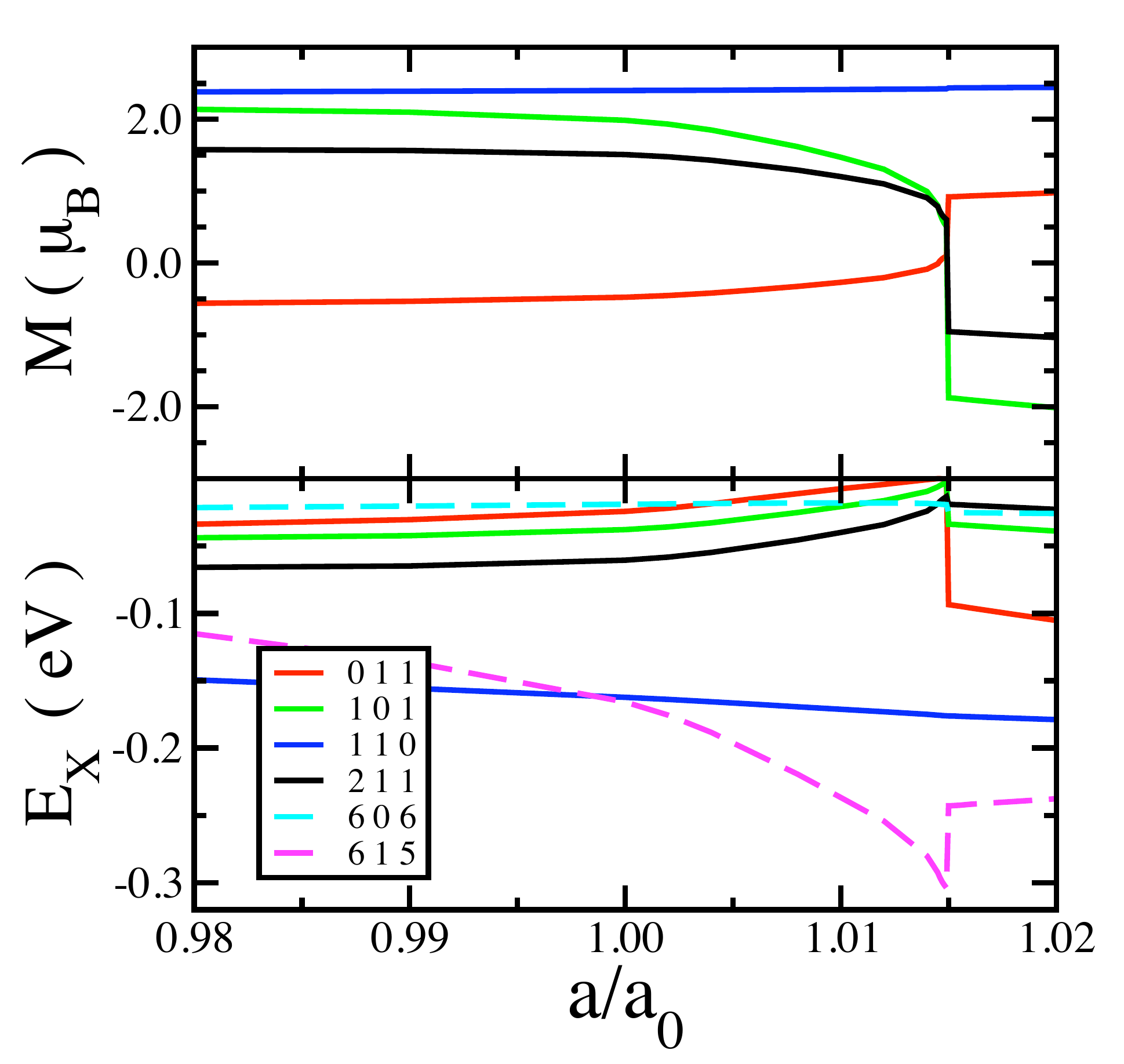}
  \caption{In the upper part, the spin, orbital and total magnetic moments, are shown as a function of variation in the in-plane lattice constant $a$ with respect to the experimental one $a_{0}$, calculated with $U=0.9$ eV.
  In the lower panel the non-vanishing contributions to the exchange energy are displayed. \label{cross}}
\end{figure}

This observation of large higher order multi-poles, naturally leads to the question: ``how relevant are they for {\urusi} with its tiny magnetic moments and hidden order parameter?''
In Fig.\ \ref{fig:exengy} we have also display the calculated moments and decomposed exchange energies for {\urusi}, set up with an antiferromagnetic order $q=(001)$. Here we see that while the overall tendency is in accordance with what we found for the other actinide compounds above, there are two details worth pointing out.
Firstly, the polarisation of the 615-channels are even more prominent and, secondly,
at large $U$ (1.2 eV) the SM switches sign and becomes parallel to the OM and both moments become very small. These properties signals some anomaly, and we proceed to study this phase under pressure.
From careful analysis of the results of experiments under uni-axial stress, one has concluded that the main variation in the pressure experiments arise from the contraction of the tetragonal $a$-axis\cite{Yokoyama:2005p4255}.
Therefore we have performed calculations with varying lattice constant $a$, to mimic the effect of pressure. The results for $U=0.9$ eV are shown in Fig.~\ref{cross}, where we see a dramatic effect on the magnetic moments from small variations in lattice constant $a$ away from a critical value, slightly (1.4 \%) larger than the experimental value $a_{0}$. At this critical value both the SM and OM vanish, while simultaneously the $615$ contribution to the exchange energy almost diverges. In fact all the vector contributions; $011$, $101$ and $211$, go to zero at this point. 
The magnetic moments only disappear for a narrow range of the in-plane lattice constant in the present study where only this parameter is allowed to vary. Future studies have to clarify whether this range increases when proper total energy optimisations, including proper Si site relaxation as well as variation of the $c$-axis, are taken into account. 

\begin{figure}
  \centering
  \includegraphics[width=\columnwidth]{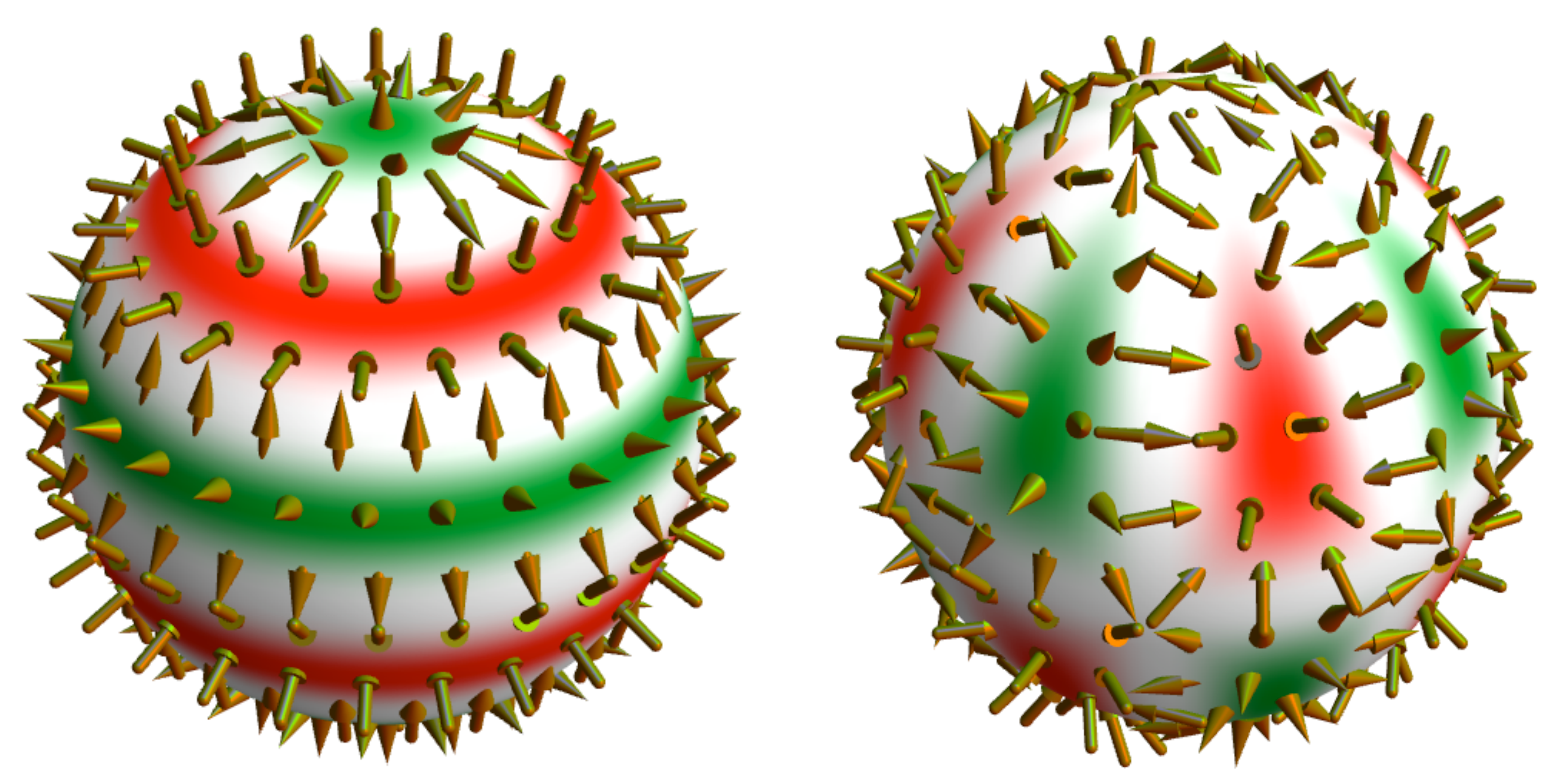}
  \caption{The angular variation of the direction of the spin density for the two non-vanishing components of the triakontadipole, ${w}^{615}_{0}$ and ${w}^{615}_{\pm4}$. The green (red) colour indicates the regions where the spin axis is outward (inward) normal.
  }
  \label{m_615}
\end{figure}

These results are very sensitive to the Brillouin zone integrations, which indicates that the Fermi surface nesting\cite{Elgazzar:2008p2090} of the un-polarised state play a major role in this instability. This nesting will lead to a large degeneracy that can be split by TR breaking polarisations.
Since in the present case the effective spin-orbit coupling is very large this TR breaking cannot be of usual spin-splitting. 
Instead it is more convenient to study this TR breaking in the $j=5/2$ subset of the $f$ shell. Then it is possible to see that it is order parameters of the tensor type $\mathbf{w}^{k1(k-1)}$ that occurs after the symmetry breaking, with larger contribution the larger $k$. This is why the $615$ tensors moments enter. 
Further analysis gives that there are two components of the $615$ tensor that are allowed by symmetry. Both are large but it is the $w^{615}_{\pm4}=(w^{615}_{4}+w^{615}_{-4})/\sqrt{2}$ component that drives the rapid variation in exchange energy as function of the in-plane lattice constant, while $w^{615}_{0}$ are almost constant. 
The spatial variation of the magnetization density arising from these triakontadipole components are displayed in Fig.~\ref{m_615}. From this plot we can note that it is closely related to the intra-atomic non-collinear spin density that always arise in the presence of spin-orbit coupling \cite{Nordstrom:1996p1264}. Since the exchange coupling enhances the effective spin-orbit coupling substantially, the resulting $615$-polarisation becomes more important than the ordinary spin-polarisation $011$. For all cases except UAs its exchange energy contribution bypass the one of spin-polarisation for values of $U$ around 1 eV.
This casts, together with studies on other magnetic actinides\cite{Long}, some new light on the magnetism of uranium systems and actinides in general.

From Fig.~\ref{cross} we see that for smaller lattice constant, $a$, antiferro-magnetic moments rapidly occur. For low pressures the SM takes a value of about 0.5 $\mu_\mathrm{B}$, while the OM is larger and of opposite sign. This state is in a sense anomalous too since the SM is now anti-parallel to the $w^{615}_{0}$ moment, while 
they are parallel in the limit of weak spin-orbit coupling. However as is clear from our calculations, this AF order is of minor importance since it is still the HO, the $615$ multi-pole, that dominates the exchange energy.
Although the fact that the AF state stabilised under pressure is closely connected to the HO phase in line with the concept of adiabatic continuity\cite{Jo:2007p4183}, there are two different solutions in the calculation and the transition from one to the other sensitively depends on the value of $U$ in our calculations. For values of $U$ above 1.2 eV the transition is clearly of second order, while for values below 0.8 eV it is of first order. In between these values the numerical accuracy is not enough to safely determine the order. However, in order to properly describe this phase transition, accompanying lattice relaxations have again to be fully taken into account, which is out of scope of the present study.

Can the 615-tensor order parameter be observed at this critical lattice point where dipole tensors vanish? 
Well, due to its high rank it is indeed a well hidden order parameter, but since there is a magnetization density associated with it, as in Fig.~\ref{m_615}, it will give rise to magnetic scattering in e.g.\ neutron diffraction (ND) experiments although the integrated moment is zero. 
One problem though is that it belongs to the same point group representation as any non-vanishing dipole order\cite{Walker:1993p191}, so very careful analysis on high accuracy experiments is needed to distinguish this pure HO case from a tiny dipole moment case. 
However, if it is this triakontadipole-state which is the HO, it can resolve a puzzling discrepancy between the tiny moments observed in ND\cite{Broholm:1987p6230,Walker:1993p191} with the lack of local spin splitting detected by  e.g.~NMR experiments\cite{Matsuda:2001p5237}: It is not tiny dipole moments that give rise to the spin-flip observed in the ND experiments, but the non-collinear 615-multipoles, which by nature have an extremely short range stray field and hence leads to no hyperfine-fields at the probing sites.
Hence, there would be no need to invoke the idea of inhomogeneous ordering\cite{Matsuda:2001p5237}, i.e.~that AF order exists in part of the sample while HO exists in the rest.


In the calculations of the magnetic uranium compounds we generally find a coexistence of polarisations of the 615 channel and the 011 channel. 
An analysis of the properties of the density matrix gives that a 011 polarisation is always permitted, and usually favourable, whenever full saturation of the other channels have not been reached.
However, when both the ${w}^{110}_{0}$ and ${w}^{615}_{0}$ tensor components are saturated,
an additional saturation of ${w}^{615}_{\pm4}$ leads to that the density matrix would acquire negative eigen-values by any spin polarisation ${w}^{011}_{0}$.
Hence, these two latter types of polarisation are competing, in accordance with our calculated results for {\urusi} in Fig.~\ref{cross}.
In this respect, the observation of a HO of {\urusi} is the effect of that the polarisation of $w^{615}_{\pm 4}$ is nearly optimal and hence forbids the usual spin polarisation.

In conclusions, we find that the  triakontadipoles $\mathbf{w}^{615}$ play a major role in all magnetic light actinides. What is unique in the case of {\urusi} is that its polarisation is so large that the usual dipole polarisations, e.g.\ spin polarisation, are forced to vanish. Under pressure, the 615 polarisation decreases and dipole polarisations are again allowed. In experiments this manifests as an apparent dipole AF order, although we find in our calculations that the HO candidate 
$\mathbf{w}^{615}$ still dominates the physics of the 5$f$ shell. Our results imply that there is always an hidden order in the magnetic actinides, it is only when it forces the dipoles to vanish as in {\urusi}  
it becomes perceptible. 

The support from the Swedish Research Council (VR) is thankfully acknowledged. The computer calculations have been performed at the Swedish high performance centers HPC2N 
under grants provided by the Swedish National Infrastructure for Computing (SNIC).

\end{document}